\documentclass[acmlarge]{acmart}

\usepackage{booktabs} 
\usepackage{tabto}
\usepackage[first=0,last=9]{lcg}

\usepackage{subfigure}

\usepackage{amsmath}
\usepackage{amsfonts}

\usepackage{multirow}
\usepackage{textcomp}
\usepackage[normalem]{ulem} 
\usepackage{soul} 

\usepackage[ruled]{algorithm2e} 

\SetAlFnt{\small}
\SetAlCapFnt{\small}
\SetAlCapNameFnt{\small}
\SetAlCapHSkip{0pt}
\IncMargin{-\parindent}

\setcopyright{usgovmixed}

\acmDOI{0000}

\received{February 2007}
\received{March 2009}
\received[accepted]{June 2009}

\usepackage{tikz}

\newcommand{\sdk}{AFV}
\begin{document}
\title{Virtualized Application Function Chaining: Maximizing the Wearable System Lifetime} 

\author{Harini Kolamunna}
\orcid{0000-0002-0761-529X}
\affiliation{%
  \institution{University of New South Wales and Data61-CSIRO}
  \country{Australia}}

\author{Kanchana Thilakarathna}
\affiliation{%
  \institution{University of Sydney and Data61-CSIRO}
  \country{Australia}
}
\author{Aruna Seneviratne} 
\affiliation{%
 \institution{University of New South Wales and Data61-CSIRO}
 \country{Australia}}

\begin{abstract}
The number of smart devices wear and carry by users is growing rapidly which is driven by innovative new smart wearables and interesting service offerings. This has led to applications that utilize multiple devices around the body to provide immersive environments such as mixed reality. These applications rely on a number of different types of functions such as sensing, communication and various types of processing, that require considerable resources. Thus one of the major challenges in supporting of these applications is dependent on the battery lifetime of the devices that provide the necessary functionality. The battery lifetime can be extended by either incorporating a battery with larger capacity and/or by utilizing the available resources efficiently. However, the increases in battery capacity are not keeping up with the demand and larger batteries add to both the weight and size of the device. Thus, the focus of this paper is to improve the battery efficiency through intelligent resources utilization. We show that, when the same resource is available on multiple devices that form part of the wearable system, and or is in close proximity, it is possible consider them as a resource pool and further utilize them intelligently to improve the system lifetime. Specifically, we formulate the function allocation algorithm as a Mixed Integer Linear Programming (MILP) optimization problem and propose an efficient heuristic solution. The experimental data driven simulation results show that approximately 40-50\% system battery life improvement can be achieved with proper function allocation and orchestration.

\end{abstract}

\maketitle

\renewcommand{\shortauthors}{H. Kolamunna et al.}

\section{Introduction}

Interest and popularity of mobile and wearable devices is persistent and is increasing due to the variety of attractive applications and services provided on top of them. Mixed Reality is expected to drive the future demand as they utilize advanced capabilities of the smart wearables and hand-held devices that allows users to interact with immersive virtual environments with real objects \cite{Costanza:2009:MRS:1532769.1532773}. On the other hand, many novel services including mixed reality are extremely resource hungry as much as revolutionary. 

These applications continuously use three types of functions, namely sensing, processing and communication. Each function in these broader categories can be considered as `application functions', as they are utilized during the runtime of the applications. These application functions requires a considerable amount of system resources such as energy, computation, and memory. Among these, energy has been identified as the most challenging, as the battery technology is not keeping pace with growth in demand. As a result,  the life time of current resource heavy wearable applications that rely on wearable devices cannot be used for long periods of time  \cite{7058994, 7993011}.


Today, most users own more than one wearable device. According to Cisco, average devices and connections per capita in North America will rise to 12.9 by 2021 \cite{cisco17}. These devices can be classified as either a Tier 1 or a Tier 2 device depending on their functional capabilities that also reflects the  resources they posses \cite{Kolamunna2016}. Tier 1 devices, e.g. smartphone, smartwatch and smartglasses, are more resourceful than the Tier 2 devices and have the capability of performing all three types of the functions that many applications require, namely sensing, processing, and communication (short-range and long-range). On the other hand, Tier 2 devices, e.g. heart-rate sensor, smart-sole, etc. basically perform  sensing functionality and has short-range communication. They are generally paired with a Tier 1 device. These devices are already interconnected and form a personal area network on user's body (PAN). Many devices on a PAN replicate the three types of functions mentioned above. Therefore, as the devices in a PAN are interconnected, it is possible to consider them as providing a pool of functional capabilities that are distributed and often replicated in multiple of the devices. 

This open us the possibility of utilizing capabilities across devices on a PAN. However,  the current system implementations limits the accessibility of  resources on the other devices.  Methods of overcoming these limitations and   allowing seamless access to the resources on a PAN and  in  close proximity has been proposed \cite{Kolamunna2016, Vallina-Rodriguez:2011, Hemminki2013, MobilePlus}. These approaches allow the functions to be implemented and the user/application to access them without considering the exact location of the function. To fully leverage the distributed resources,  the sequences of functions (function chains) execution needs to be guaranteed, to achieve the correct \textit{Functionality}. Therefore, the challenge is to determine the criteria for the selection of the path of the function chain, and how does this dynamic function chaining can be implemented where the functions are virtualised to enable seamless accessibility similar to network function orchestration in software defined networks \cite{Bari, 6984061}. 
This requires the design of an optimal  application function orchestration algorithms. This paper shows that optimal application function orchestration is NP-hard, but that it possible to achieve near optimal function orchestration using a heuristic algorithm that finds the best function allocation by considering the function chaining in order to maximize the system lifetime, and make the following contributions. 

\begin{itemize}
	\item Define design goals for the wearable system energy resource management considering the current trends in wearable devices as well as surveying diverse set of users to understand user expectations. 
	\item Formulate the function allocation problem in wearable systems as a Mixed Integer Liner Programming (MILP) optimization to maximize the wearable system lifetime whilst satisfying all functional requirements.
	\item Derive the linear programming approximation factor and show the requirement of efficient heuristic algorithm under practical constraints, which is followed by an efficient heuristic algorithm incorporating function orchestration.
	\item Demonstrate the feasibility of implementing the proposed heuristic algorithm on real devices by measuring and comparing the computational time on a smartphone and a smartwatch.
	\item Evaluate the proposed heuristic solution accuracy, efficiency, and robustness, against the currently used function allocation methods and methods proposed in the literature by conducting experimental data driven simulations. The results show that dynamic function allocation increases the system lifetime by approximately 50\% compared to the existing common non-collaborative function allocation methods and  40\% compared to the collaborative function allocation methods proposed in the literature. 
\end{itemize}

%
The rest of this paper is organized as follows. 
We first overview the related work and complementary systems. Next in Section 3, we present the function allocation problem formulation followed by efficient algorithms for function allocation including function orchestration in Section 4. In Section 5,  we evaluate the proposed algorithms and also the feasibility of increasing wearable system lifetime with experimental driven simulations. Finally, we provide the conclusions in Section 6. 

\section{Related Work}
\label{related_work}

The concept of application function virtualization (AFV) is an extension and generalization of the concept of  network function virtualisation (NFV). As in the case with NFV, one of the challenges is the the chaining of the functions (service function chaining - SFC) to achieve the dual objectives of providing the required services, and maximizing the service lifetime. 

NFV enables the functions that are provided by proprietary hardware and software to being able run on open hardware, thus reducing the capital and operational expenditure. However, to provide specific network service in an NFV enabled network, requires the traffic to traverse multiple of the virtualized function instances in a defined order. Unlike in the hardware implementation of the network functions, the virtualization of the functions makes this too complex and inefficient \cite{7524565}. More specifically, the virtualized functions are topology dependent and  difficulties occurs during the configuration \cite{IETF}. Therefore, much attention has been paid to developing mechanisms for chaining virtualized functions or orchestration by both the research community and standardization authorities \cite{IETF2,Bari}. 

In order to perform function chaining, it is required to have knowledge about the available services via the virtualized function instances and their reachability. There are several approaches available for SFC that extends the currently available methods  \cite{6702549,6391820}. The architectural designs proposed by the standardizing authorities \cite{ETSI2014, IETF2} build on top of the basic NFV architectures proposed by \cite{ETSI2014}. Similarly, in our design, we build our function chaining architecture on top of the proposed application function virtualization architecture (\sdk) in \cite{Kolamunna2016}  and extend it to incorporate function chaining. These architectural designs for function chaining have the following approaches; 1) having a separate management system to orchestrate the function chaining \cite{ETSI2014,Li2015}, which allows the control plane to have the overview of the availability of function instances and how they can be accessed, and 2) allowing the data packets to carry the controlling signals within its header \cite{IETF2,ietf-sfc-nsh-28}. 

Our design combines both the approaches but generalizes it to be used with application function virtualization requirements, which leads to  different objectives such as energy minimization \cite{6461195}, optimal utilization of the resources \cite{SAHHAF2015492, 6876564, 6968961}, minimization of latency \cite{BHAMARE201568,7014205}, minimizing the monetary cost \cite{7116121}, etc.  None of these proposed schemes consider application function virtulalization and attempts to combine the different objectives to maximize the system lifetime.

\section{Problem Formulation}
\label{sec:FAP}

 \begin{figure}[tb]
 	\centering 
 	\includegraphics[width=0.8\textwidth]{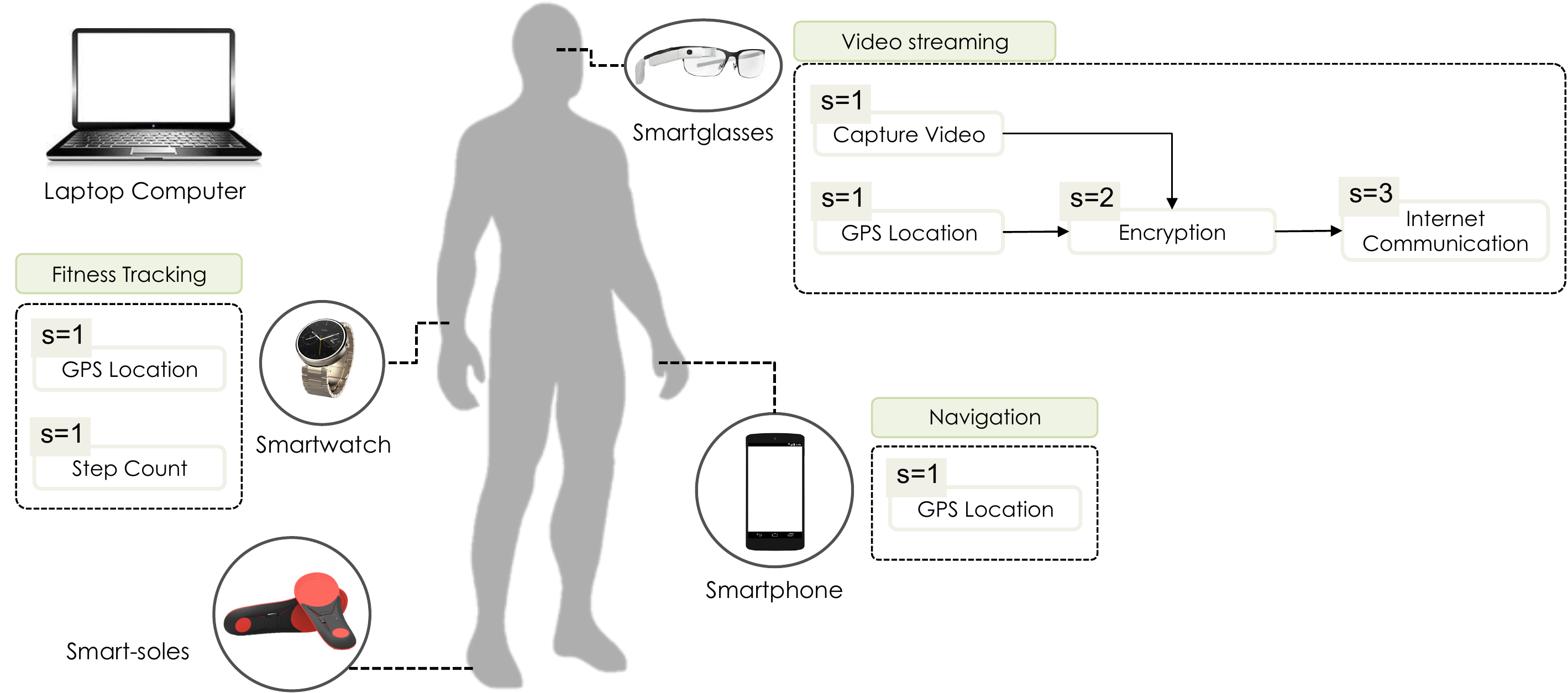}
 	\caption{An overview of an personal area network.}
 	\label{fig:pwn}
 \end{figure}

Despite the increasing trend in wearing and carrying multiple personal devices, there is an significant overlap in primary functionality they provide such as sensing, processing and communication, especially among the Tier 1 devices. In the example PAN in Figure \ref{fig:pwn}, all three Tier 1 devices (i.e., smartwatch, smartphone and smartglasses) can perform GPS location sensing, step counting and provide Internet access. However, each device is likely to have specific use cases (i.e., fitness tracking, navigation and video streaming), which combine and use these primary functions in different orders. Our hypothesis is that \emph{if the devices on the same PAN collaborate with each other to harness the power of common functions, it is possible to provide better utility to the user}. In this paper, we focus on the \emph{wearable system lifetime} as the utility objective, although it can be other objectives such as new services, information quality and/or information fidelity/precision. 

Wearable system lifetime can be defined in multiple ways as per individual user requirements: i) All devices to be useable for the longest period of time,  ii) At least one device to be useable and iii) The preferred device to be useable for the longest period of time. Therefore, we first conducted a user study to understand user requirements in terms of  lifetime of devices on a PAN. 

\subsection{User study: Wearable system lifetime}


We collected responses from 70 users who uses more than one wearable or a hand-held device. These users belonged to different age groups (20-60 years) and  come from different backgrounds (e.g. Healthcare, Education, Technology, Student, Sales/Service, etc.).

 \begin{figure}[tb]
 	\centering 
 	\includegraphics[width=0.75\textwidth]{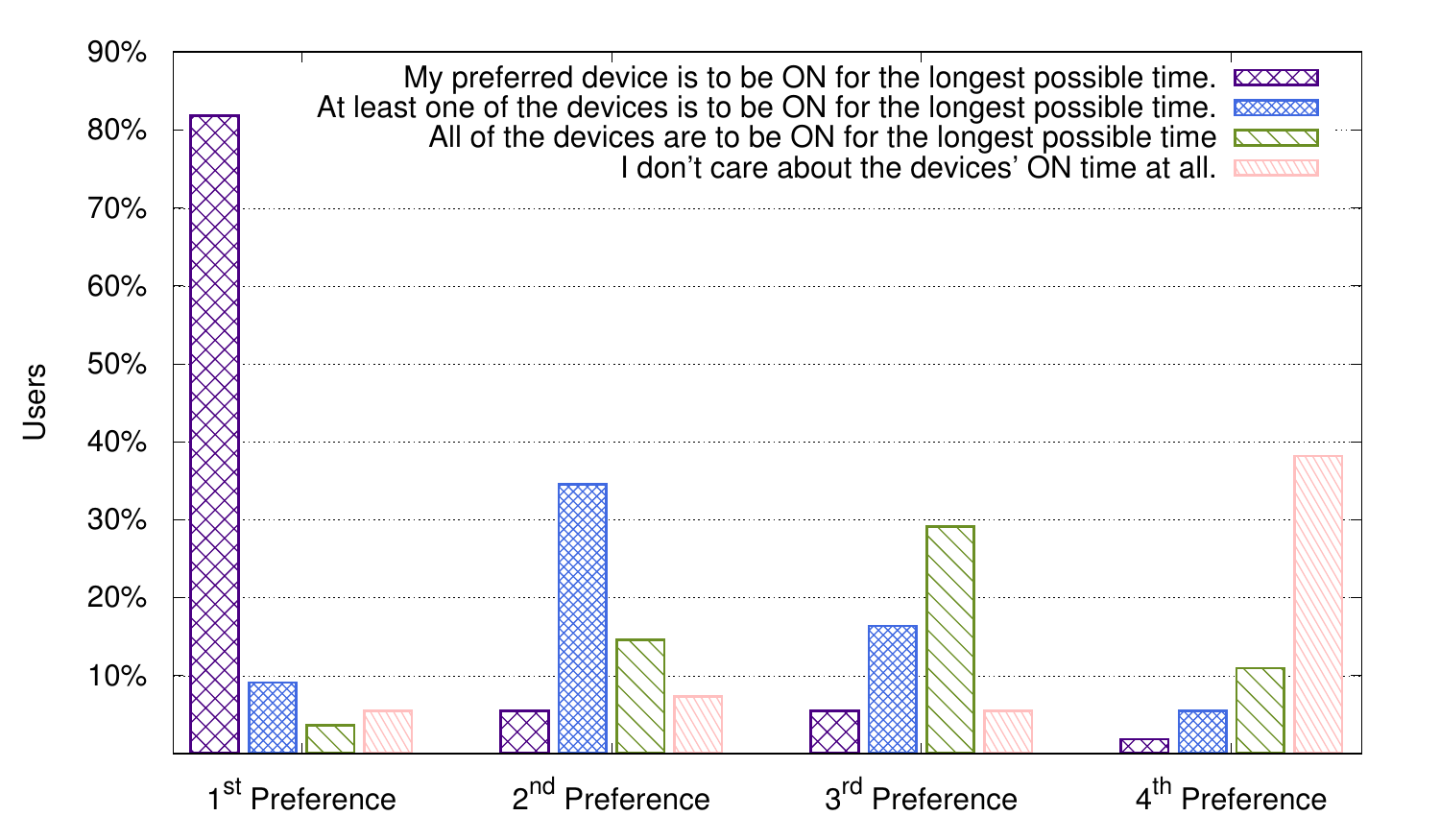}
 	\caption{User preference in wearable system lifetime.}
 	\label{fig:survey}
 \end{figure}

According to our survey results (Figure \ref{fig:survey}), overwhelming majority voted for the preferred device to be useable for the longest period of time. Further analyses of user comments showed that it is more about their preferred functionality, instead of their preferred device.  
However, the availability of the user preferred functionalities are replicated in todays Tier 1 devices, and even some applications are designed to incorporate the functionalities of other devices as well (e.g. mixed reality applications). 
In light of these user responses, we define the wearable system lifetime as \emph{the battery life of Tier 1 devices on the PAN}. This is followed by our problem statement \emph{how to maximise the battery life of Tier 1 devices whilst satisfying all functional requirements}.

\subsection{Function Allocation Problem}

Thus the objective is to allocate each function request $r^i_f \in R_i$ from each device $i \in D$ to function implementations $v_f \in V_i$ in different devices in the correct order whilst maximizing the  system lifetime. In other words, for each instance, it is required to maximize the lifetime of the Tier 1 device on the PAN with the lowest battery life. We define the lifetime of a device as the ratio between the remaining energy ($E_i$) and expected energy usage which is defined as the sum of current usage ($C_i$) and the energy usage of newly assigned functions ($A_i$). We then formulate the function allocation problem as a \emph{Mixed Integer Liner Programming (MILP)} problem in Equation \ref{eq:test1}. Table \ref{tbl:notation} summarises the definitions for all the symbols used in this section. 


%
%
%
%

\begin{equation}\label{eq:test1}
\begin{aligned}
{Maximize}  \Bigg\{\textrm{\textit{Min of}} \bigg\{\bigg[ \frac{E_i}{(C_i+ A_i)} \bigg]\forall i \in D\bigg\}\Bigg\} \textrm{~~~~~~~~~~~~~~OR~~~~~~~~~~~~~~} {Minimize}  \Bigg\{\textrm{\textit{Max of}} \bigg\{\bigg[ \frac{(C_i+ A_i)}{E_i} \bigg]\forall i \in D\bigg\}\Bigg\}
\quad \text{where} \\
A_i  =  \left(\sum_{v^i_f \in V_i}f_{v^i_f}\cdot  w_{v^i_f} + \sum_{j \in D} \sum_{r^i_f\in R_i} c_{j,r^i_f} \cdot x_{j,r^i_f}+ \sum_{j \in D} \sum_{v^i_f\in V_i} c_{j,v^i_f} \cdot y_{j,v^i_f} + H_i\cdot z_i\right)
\end{aligned}
\end{equation}


%
\noindent Such that:
\begin{flalign*}
	1. \quad & w_{v^i_f} , x_{j,r^i_f}, y_{j,v^i_f}, z_i \in \{0,1\}; \hspace{0.4cm} {\forall (r^i_f \in R_i), \forall (v^i_f \in V_i), \forall (i,j \in D)}\\
	2. \quad & c_{j,r^i_f} , c_{j,v^i_f} = 0 ;\hspace{0.4cm} \textrm{if} \hspace{0.4cm} j=i\\
	3. \quad & x_{j,r^i_f} =\begin{cases}
		1;& \text{if \hspace{0.4cm}$v^j_f \in V_j$}; \hspace{0.4cm}\forall (r^i_f \in R_i), \forall (i \in D) \\
		0;& \text{else}
	\end{cases}\\
	4. \quad &\sum_{j \in D} x_{j,r^i_f} =1; \hspace{0.4cm} \forall (r^i_f \in R_i), \forall (i \in D)\\
	5. \quad & x_{j,r^i_f}=y_{i,v^j_f}\\
	6. \quad & w_{v^i_f} =\begin{cases}
		1;& \text{if \hspace{0.4cm}$\sum_{j \in D} y_{j,v^i_f}\geq 1$}; \hspace{0.4cm}\forall (v^i_f \in V_i), \forall (i \in D) \\
		0;& \text{else}
	\end{cases}\\
	7. \quad &z_{i} =\begin{cases}
		1;& \text{if \hspace{0.4cm}$\sum_{j \in D; j\ne i} \sum_{r^i_f\in R_i}x_{j,r^i_f}\geq 1$ or $\sum_{j \in D; j\ne i} \sum_{v_f\in V_i} y_{j,v_f}\geq 1$} \\
		0;& \text{else}
	\end{cases}\\
	8. \quad &\bigg[ \frac{C_i}{E_i}+ \frac{1}{E_i}*\left(\sum_{v^i_f \in V_i}f_{v^i_f}\cdot  w_{v^i_f} + \sum_{j \in D} \sum_{r^i_f \in R_i} c_{j,r^i_f} \cdot x_{j,r^i_f}+ \sum_{j \in D} \sum_{v^i_f\in V_i} c_{j,v^i_f} \cdot y_{j,v^i_f} + H_i\cdot z_i\right) \bigg]\leq \frac{1}{T_i^*};\hspace{0.1cm}\forall (i \in D)\\
\end{flalign*}

$A_i$ depends on the energy consumption of each function, \emph{Function Cost} $f_{v^i_f}$ and the energy consumption for the data transfer in between two devices (\textit{Communication Cost}), if the function is allocated to another device. Communication cost consists of two parts, i.e. \textit{data transferring cost}, and \textit{idle cost}. The data transferring cost can be either due to data transmission from/to a function invocation in device $j$ to the request $r^f$ in device $i$ ($c_{j,r^i_f}$), or the data transmission from/to the function $v_f$ in device $i$ to the device $j$ ($c_{j,v^i_f}$). The \textit{idle cost} in device $i$ ($H_i$) is due to the waiting time before closing the connection. 

The first constraint shows that the execution of a function or an assignment of a function request can only be \textit{`yes'} or \textit{`no'} [0/1].
Specifically, $w_{v^i_f}=1$ if the function $v_f$ is executing in device $i$, $x_{j,r^i_f}=1$ if the request for the function $v_f$ from device $i$ ($r^i_f \in R_i$) is assigned to device $j$, $y_{j,v^i_f}=1$ if the request for the virtual function $v_f$ from device $j$ is assigned to device $i$, and $z_i=1$ if the idle communication energy is considered. Secondly, the communication costs are zero if the function invocation device and the requesting device are the same. Third constraint expresses that the request $r_f$ is mapped to device $j$ only if the function $v_f$ is available in the device $j$. The forth constraint expresses that each of the requests is assigned exactly to one of the devices that has the function. Next constraint shows that if the communication cost is considered for a particular communication in between device $i$ and $j$, both the devices need to consider the costs. The constraint 6 shows that a function will invoke only if any of the requests are assigned to the function in the device. Next, if any communication is happening, the idle communication is to be considered. Finally, if there are any requirements for any of the devices to be alive for a certain time, that requirement is to be satisfied.

In this problem formulation, the objective function and the constraints always have a linear relationship with each of its variables. However, some variables in this problem formulation has to have binary values. As an example, the assignment of a function to a request is either \textit{true(1)} or \textit{false(0)}. 
Therefore, this problem formulation is equivalent to \emph{Mixed Integer Liner Programming (MILP)} problem.

\begin{table}
\caption{Definitions of Notations}
	\centering
	\begin{tabular}{lp{13cm}}
	\hline
	Symbol & Definition\\ \hline
	{$D$} &{- Set of devices in PAN}\\
	{$V_i$} &{- Set of 
		functions in device $i \in D$}\\
	{$R_i$} &{- Set of requests made from device $i \in D$}\\
	{$T$} &{- Interval in between two executions of the algorithm}\\
	{$f_{v^i_f}$} &{- Implementation energy of a 
		function $v_f \in V_i$ per $T$ time units}\\
	{$c_{j,r^i_f}$} &{- Energy of per $T$ time units communication for the request for the function $v_f$ from device $i$ ($r^i_f \in R_i$) that is made to device $j$ (receiving energy excluding the idle cost). $c_{j,r^i_f} = 0$ if $j=i$}\\
	{$c_{j,v^i_f}$} &{- Energy of per $T$ time units communication for the request made from device $j$ to the virtual function $v_f \in V_i$ (transmission energy excluding the idle cost). $c_{j,v^i_f} = 0$ if $j=i$}\\
	{$H_i$} &{- Idle energy for device $i \in D$}\\
	{$T_i$} &{- lifetime of $i \in D$ in $T$ time units}\\
	{$T_i^*$} &{- Expected minimum lifetime of $i \in D$ in $T$ time units}\\
	{$E_i$} &{- Remaining energy of $i \in D$}\\
	{$U_i$} &{- Energy usage of $i \in D$ with in $T$ time units}\\
	{$C_i$} &{- Current energy usage of $i \in D$ with in $T$ time units}\\
	{$A_i$} &{- Assigning energy usage of $i \in D$ with in $T$ time units}\\ \hline
\end{tabular}
\label{tbl:notation}
\end{table}

\section{Function Allocation Algorithm}

Solutions to MILP problems are  known to be \textit{NP-hard} \cite{Grotschel2006} and  the potential approaches can be broadly categorised into: i) \emph{Exact Algorithms} that guarantee to find an optimal solution, but may take an exponential number of iterations; ii) \emph{Approximation Algorithms} that provide sub optimal solutions in polynomial time, and provides a bound on the degree of sub optimality; and iii) \emph{Heuristic Algorithms} that provide sub optimal solutions, but do not guarantee the quality of the solutions, and do not guarantee the polynomial time solutions.


\subsection{Exact Algorithms} Even though the exact methods assure the optimal solutions, the complexity grows exponentially. However, these methods in the case of AFV chaining can be used where the number of devices and the functions are very low. The mostly used two exact solution methods (\textit{brute force search} and \textit{branch and bound \cite{Dupont2008}}) are shown in the Appendix \ref{BFS} and Appendix \ref{BB} respectively, with the specific solutions  and their respective computational complexities.

\subsection{Linear Programming (LP) Approximations}
However, when the number of devices and available functions  increase, the use of exact methods become impractical due to the increase in computational complexity. Therefore, the general practice is to use approximation methods. Most of these methods are based on linear programming.
In these methods, the integer constrains in the original MILP problem are replaced with the linear constrains to solve the LP problem. With these methods,  optimal solutions can be obtained in polynomial time and lead to the \textit{lower bound} of the original MILP problem solutions. If any variable that is to be an integer variable in the original problem is a non-integer in the LP relaxation solutions, these values are rounded to the nearest integer. The rounding in these \textit{integral solutions} lead to sub-optimal solutions. A greedy method is required to find the \textit{integral solutions} without significantly increasing the cost. The \textit{approximation factor} is defined as the ratio in between \textit{integral solution} and the \textit{lower bound}. 
%
For the LP representation, the constraints of the function allocation problem defined in Equation \ref{eq:test1} should be changed as follows.

\begin{flalign*}
	1. \quad &\sum_{j \in D} x_{j,r^i_f} \geq 1; \hspace{0.4cm} \forall (r^i_f \in R_i), \forall (i \in D)\\
	2. \quad & w_{v^i_f} \geq x_{i,r^i_f}; \hspace{0.4cm}\forall (v^i_f \in V_i), \forall (i \in D)\\
	3. \quad & w_{v^i_f} \geq y_{j,v^i_f}; \hspace{0.4cm}\forall (v^i_f \in V_i), \forall (i,j \in D), i\ne j \\
	4. \quad & y_{i,v^j_f}\geq x_{j,r^i_f}\\
	5. \quad & z_i \geq x_{j,r^i_f}; \hspace{0.4cm}\forall (v^i_f \in V_i), \forall (i \in D), j\ne i \\
	6. \quad & z_i \geq y_{j,v^i_f}; \hspace{0.4cm}\forall (v_f \in V_i), \forall (i \in D), j\ne i \\
	7. \quad &\bigg[ \frac{C_i}{E_i}+ \frac{1}{E_i}*\left(\sum_{v^i_f \in V_i}f_{v^i_f}\cdot  w_{v^i_f} + \sum_{j \in D} \sum_{r^i_f \in R_i} c_{j,r^i_f} \cdot x_{j,r^i_f}+ \sum_{j \in D} \sum_{v^i_f\in V_i} c_{j,v^i_f} \cdot y_{j,v^i_f} + H_i\cdot z_i\right) \bigg]\leq \frac{1}{T_i^*};\hspace{0.1cm}\forall (i \in D)\\
\end{flalign*}

From the solutions to the LP approximation, we get a solution for the system lifetime, i.e. the lifetime of the device $d$ that is the minimum among the other devices. 
%
The optimal solutions for LP approximation ($OPT(LP)$) is called the lower bound of the solutions to MILP problem. The equation \ref{eq:test7} shows the lower bound calculation.

\begin{equation}\label{eq:test7}
OPT(LP) =\frac{C_d}{E_d}+ \frac{1}{E_d}*\left(\sum_{v^d_f \in V_d}f_{v^d_f}\cdot  w_{v^d_f} + \sum_{j \in D} \sum_{r^d_f \in R_d} c_{j,r^d_f} \cdot x_{j,r^d_f}+ \sum_{j \in D} \sum_{v^d_f\in V_d} c_{j,v^d_f} \cdot y_{j,v^d_f} + H_d\cdot z_d\right) 
\end{equation}


\noindent \textbf{Integral Solution:} However, the optimal solution to the LP approximation does not guarantee a integral solution. Therefore, we need to obtain the integral solution from the $OPT(LP)$ solution. Greedy methods or certain rules sets leads to the integral solution. The below mentioned are the rules that we define in order to find the integral solution.
\begin{itemize}
	\item If any request from $d$ is not 100\% assigned to any one of the other devices in the $OPT(LP)$ solutions, it should be assigned to $d$ itself.
	\begin{itemize}
		\item []If $x_{d,r^d_f} >0 $
		\item []$x_{d,r^d_f}=1$ \& $f_{v^d_f}=1$ \& $x_{j,r^d_f}=0$; $\forall j\ne d$
	\end{itemize}
	\item If any percentage of any of the requests from other devices is assigned to $d$, that means the requests can not be fulfilled by any of the other devices. Because, if it is possible, these will not be added to the device $d$ that increases the cost. 
	\begin{itemize}
		\item []If $y_{j,r^d_f} >0$;  $\forall j\ne d$
		\item []$y_{j,r^d_f}=1$ \& $f_{v^d_f}=1$
	\end{itemize}
\end{itemize}

In the worst case, all the requests from the device $d$ is to be fulfilled by the device $d$ itself. In addition, all the requests from other devices that has the [self-functioning cost $>$ communication cost] is to be run in the device $d$.
Hence, the cost for the worst case integer solution ($INT_{sol}^{W}$) is calculated as in equation \ref{eq:test13}. 
\begin{equation}\label{eq:test13}
INT_{sol}^W =\frac{C_d}{E_d}+ \frac{1}{E_d}*\left(\sum_{v^d_f \in V_d}f_{v^d_f} + \sum_{j \in D} \sum_{v^d_f\in V_d} c_{j,v^d_f} + H_d\right) 
\end{equation}

In the best case for the optimal solution ($INT_{sol}^{B}$) for OPT(LP), all the requests from device $d$ would be fully allocated in a way that ensures the minimum cost to $d$. Then, none of the requests from any of the other devices are allocated to $d$. 
\begin{equation}\label{eq:test14}
INT_{sol}^B =\frac{C_d}{E_d}+ \frac{1}{E_d}*\left(\sum_{v^d_f \in V_d}f_{v^d_f}\cdot  w_{v^d_f} + \sum_{j \in D} \sum_{r^d_f \in R_d} c_{j,r^d_f} \cdot x_{j,r^d_f} + H_d\cdot z_d\right) 
\end{equation}

\noindent \textbf{Approximation Factor:} The approximation factor (AF) for a minimization problem is defined as the ratio of the cost of integral solution to the cost of LP optimal solution \cite{7990567}. We considered the worst case integral solution for the AF calculation. The AF is calculated as in equation \ref{eq:test15}. 

\begin{equation}\label{eq:test15}
AF =\frac{C_d + \sum_{v^d_f \in V_d}f_{v^d_f} + \sum_{j \in D} \sum_{v^d_f\in V_d} c_{j,v^d_f} + H_d}{C_d +\sum_{v^d_f \in V_d}f_{v^d_f}\cdot  w_{v^d_f} + \sum_{j \in D} \sum_{r^d_f \in R_d} c_{j,r^d_f} \cdot x_{j,r^d_f}+ \sum_{j \in D} \sum_{v^d_f\in V_d} c_{j,v^d_f} \cdot y_{j,v^d_f} + H_d\cdot z_d} 
\end{equation}

The AF for a minimization problem is always greater than 1. 
However, this method of AF calculation depends on the variable values and sometimes it would give a huge value. Therefore, LP approximation method is unsuccessful of dealing with hard capacities as in this problem. Also, the linear programming representation of this specific problem suffers from another aspect that is the increment of the number of variables as the number of function types and devices are increased.

\subsection{Heuristic Algorithms}
\label{sec:Heu}

As explained earlier, heuristic methods neither guarantee the solutions with polynomial time complexities, nor the optimality of the solution. Therefore, the design of the heuristic methods have to crafted to give a nearly optimal solution without increasing the computational complexity significantly. 

Table \ref{tab2} shows the energy usage for $T$ time units for all the possible requests to functionality mappings. The first column shows the available functionalities [$v_{i,x}$] (i.e., the function $x$ in device $i$). The second column shows the functionality cost of the function $v_{i,x}$. The cell entries of this column is a series of cost values, where each value in this series is the cost for each device in the network. However, only the device $i$ will have the cost in a particular series and the cost for all the other devices are zero. 

The columns from column 3 show the communication cost  for the functionality assignment for the requests $r_{i,x}$, where the request is coming from the device $i$ for the function $x$. Similar to other columns, the cell entries in each column is a series of cost values, where each value in this series is the cost of communication for each device in the network. If the requesting device and the functioning device are different, both devices have the communication cost. On the other hand, if the request is assigned to the same device, the communication cost is zero. If the devices that is neither the requesting nor the functioning device, the communication cost is again zero.
The values for $C_i$, $E_i$, $H_i$ and $T^*_i$ for each device are fed to the algorithm. The values for $E_i$ and $T_i^*$ do not change over the iterations for one solution search. However, $C_i$ and $H_i$ would change their values depending on the selection of functions and the requests during each iteration.
\begin{table}[tb]
	\caption{Energy usage for $T$ time units. (\textit{energy usage in $d_i$,...,energy usage in $d_n$})}
	\centering
	\begin{tabular}{|l||c|c|c|c|}
		
		\hline
		$v_{i,x}$&$f$ & $r_{1,1}$ & .~.~.~.&
		$r_{n,x}$ \\ 
		\hline \hline
		$v_{1,1}$  &($f_{1,[1,1]}$,...,$f_{n,[1,1]}$)& ($c_{1,[r_{1,1}],[v_{1,1}]}$,...,$c_{n,[r_{1,1}],[v_{1,1}]}$) & & \\ 
		\hline
		.  &.&.&& . \\
		.  &.&.&.~.~.~.& . \\
		.  &.&.&&.  \\
		\hline
		$v_{n,x}$  &($f_{1,[n,x]}$,...,$f_{n,[n,x]}$)&($c_{1,[r_{1,1}],[v_{n,x}]}$,...,$c_{n,[r_{1,1}],[v_{n,x}]}$)&&  \\ 
		\hline
	\end{tabular}
	\label{tab2}
\end{table}

\begin{algorithm}[tb]
	\label{algo:fap1}
	\SetAlgoNoLine
	\KwIn{$(R_v, D_v, m, f, c, T)$}
	\KwOut{assignment $\sigma: R_v \mapsto D_v$}
	$T^{*} = T^{**}$;
	
	\Repeat{$R_v \neq \emptyset$}{
		Select $v \in D_v$ and $R \subseteq R_v$ s.t. $\forall r \in R:$ that minimize
		$\frac{T^*-T}{|R|}$ 
		
		$R_v$ $\leftarrow$ $R_v-R$; $f_{v}=0$\\
		$C_i$ $\leftarrow$ $C_i+A_i$ \\
		$T=T^*$\\
		$h_i =\begin{cases}
		1;& \text{if \hspace{0.4cm}$\sum_{r \in R} c_{i,r,v} > 0$}; 
		\\
		0;& \text{else}
		\end{cases}$\\
		$H_i$ $\leftarrow$ $H_i-H_i\cdot h_i$\\
		$(R_v \mapsto D_v$) $\leftarrow$ $(R_v \mapsto D_v$) $+$ ($R$ $\mapsto$ $v$)
	}
	\caption{Function Allocation Algorithm}
	\label{alg:one}
\end{algorithm}

The algorithm runs for each type of function separately as a particular function request can only be satisfied with the invocation of the same type of function. 
The algorithm works as follows for a particular function type $x$, and described in Algorithm 1. The system lifetime is calculated as in (\ref{eq:test5}). Let's assume that $T^{**}$ is the initial system lifetime that is calculated without assigning the requests. First, we select a function $v$ ($v_{i,x}$) and a set of requests $R_v$ that minimizes the system lifetime reduction from the earlier system lifetime. And once the function and the set of requests are selected, the set of requests are assigned to the function and remove from further considerations. However, the function is still considered with the functioning cost of zero, as it is already executing for the assigned requests. Then the current energy usage $C_i$ for the devices are reassigned with the cost $(C_i+A_i)$. Then, the $H_i$ is replaced by the value $(H_i-H_i\cdot h_i)$, where $h_i$ indicates whether the device $i$ is involved in communication with any of the other devices.

\begin{equation}\label{eq:test5}
T={\textrm{\small \textit{Min of}} }  \Bigg\{\frac{1}{ \bigg[ \frac{C_i}{E_i}+ \frac{1}{E_i}*\left(\sum_{v^i_f \in V_i}f_{v^i_f}\cdot  w_{v^i_f} + \sum_{j \in D} \sum_{r^i_f \in R_i} c_{j,r^i_f} \cdot x_{j,r^i_f}+ \sum_{j \in D} \sum_{v^i_f\in V_i} c_{j,v^i_f} \cdot y_{j,v^i_f} + H_i\cdot z_i\right) \bigg]}(\forall i \in D)\Bigg\} \\
\end{equation}

\subsection{Virtualized Function Chaining in \textsc{faa}}
\label{sec:FAA}
In the previous section, we show how the the proposed heuristic method of \textsc{faa} runs for each type of functions separately and assigns these functions optimally. In this section, we show the incorporation of function chaining in \textsc{faa} in order to find the optimal path of the function chain. 

Each of the real world application has its own sequence of functions (cf. Figure \ref{fig:pwn}). In this figure, the sequence number `$s$' is labeled for each of the function in each application. One device may have multiple of the applications running at a particular time. Therefore, a particular sequence of functions $S=[s_1,s_2,...,s_n]$ is defined for each application `$a$' in device `$d$'. Therefore, each request has its unique \textit{Request ID} ($< d,a,s>$).

At first, all the types of function requests that has the sequence number $s=1$ are considered. Next, the \textsc{faa} is performed for each of the function type and finds the best allocation for each type. Then it considers the functions in the ascending order of the sequence number. In order to maintain function chaining, a log is maintained as the \textsc{faa} is performed. This log contains $<$\textit{chain identification ($d,a$), sequence number ($s$), function type ($f_i$), function requesting device ($d_r$), function performing device ($d_f$)}$>$. After performing \textsc{faa} for all function types for a particular $s$ value, the \textsc{faa} then starts allocating functions to the requests that are next in the sequence ($s+1$). At this point, the \textsc{faa} does not consider that these requests are coming from the original requesting devices, but they are considered as coming from the devices where the previous request in the chain is allocated to. 

\section{Evaluation}

We first present the performance analysis of \textsc{faa} in terms of accuracy, efficiency, and robustness. Next, we evaluate the wearable system lifetime improvement using virtualized function chaining with an use case scenario that is emulated with an experimentally measured energy consumption values. 


\subsection{Performance analysis of \textsc{faa}}

\subsubsection{Simulation setup}
For this analysis, we consider a wearable system with three devices, where each device is installed with the proposed virtualized function chaining modules. Each function chain is simulated to have the same number of functions that is gradually increased up to ten functions per device. We consider every function is of different type and the energy consumption of each function (\textit{Function Cost}) is normally distributed with average $\mu$ of 200mJ per minute, and the standard deviation of ($\sigma$) such that $\sigma = 0.1*\mu$. We consider that each function uses 13.5KB of data per minute (i.e. a typical amount of data generated from a sensing function per minute \cite{Kolamunna2016}) that needs to be transferred to the next function in the chain once in every minute. We consider that the three devices have 400mAh, 450mAh, and 500mAh battery capacities (i.e. available smart wearable devices' normal battery capacity \footnote{https://www.gsmarena.com}), and each of the experiment starts with 100\% of battery charge in all the devices.

The energy consumption for the data transfer in between two devices (\textit{Communication Cost}) consists of two parts, i.e. \textit{data transferring cost}, and \textit{idle cost} that is caused due to the waiting time before closing the connection. In general, the \textit{idle cost} is almost 80\% from the \textit{Communication Cost} during Bluetooth transmission of $\sim$10KB of data. The \textit{Communication Cost} for a function is as same as the average \textit{Function Cost} (200mJ per minute). However, we consider that all the communication in between devices (10KB of data for each function) is performed once per minute, and all at once. Therefore, the \textit{idle cost} is shared among the number of functions in the chain. Hence, as the number of functions in the function chain is increased, the \textit{Communication Cost} per function is decreased.

We then compare the system lifetime (i.e.  time until all the Tier 1 devices are 'ON') increment of (\textsc{faa}) against the following;
\begin{itemize}
\item OPTIMAL - The optimal solution for every \textsc{faa} instance obtained from the brute force method.
\item MANUAL - A random selection of one device in the wearable system to execute each function request. This method is analogous to asking the user to select function allocation.
\item EACH - Every device in the wearable system perform all functions independently which is the usual practice with current cross device wearable applications.
\item $FAA_{AFV}$ - The most related function allocation algorithm proposed in the literature \cite{Kolamunna2016}. However, this work does not consider function chaining. Therefore, we modified $FAA_{AFV}$ to incorporate function chaining as described in Section \ref{sec:FAA}.
\end{itemize}



\subsubsection{Accuracy, Efficiency and Robustness of \textsc{faa}}
\label{Eval}

\begin{figure}[t]
	\centering  
	\subfigure[The impact of no. of functions for the optimality of \textsc{faa} algorithm.]{\label{fig:optimscaling}\includegraphics[width=0.48\textwidth]{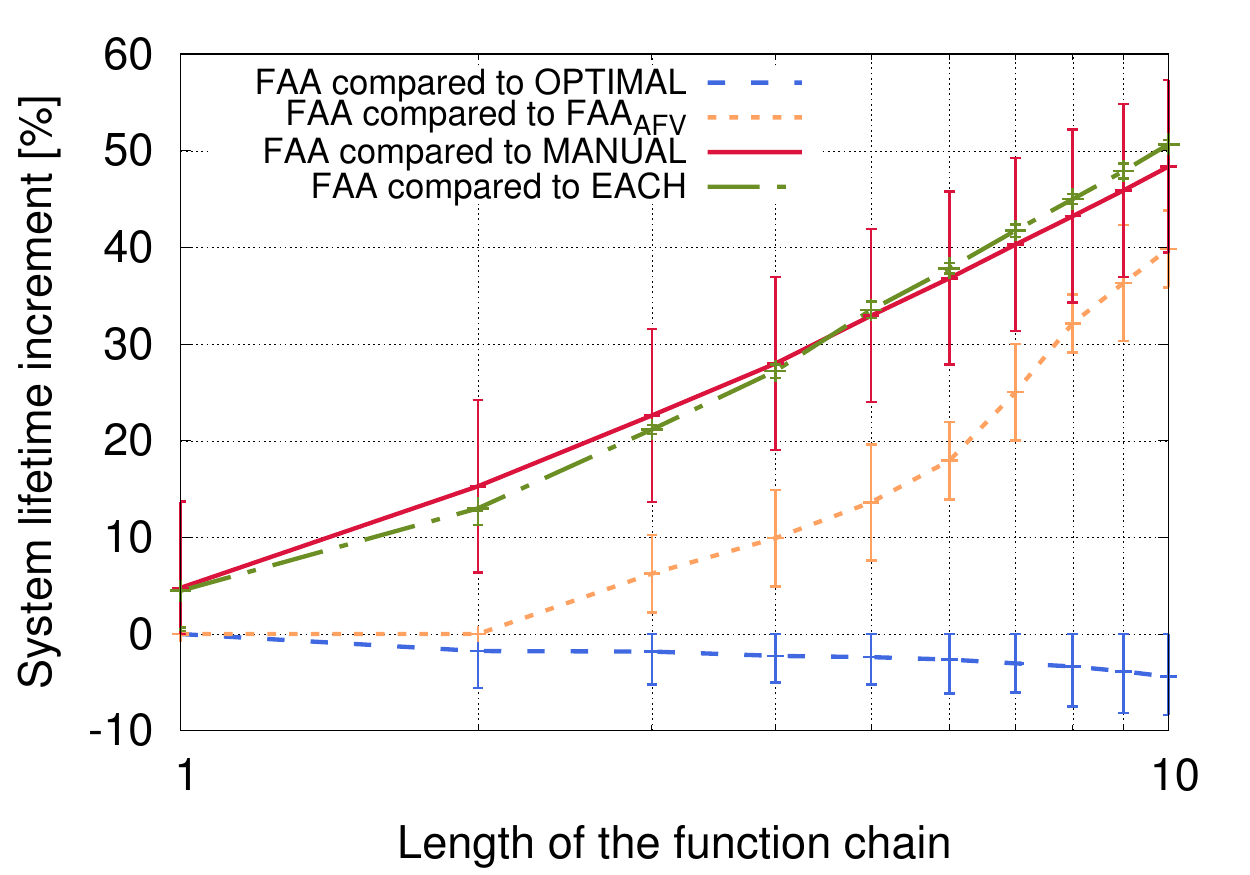}}
	\subfigure[Algorithm execution time.]{\label{execution}\includegraphics[width=0.48\textwidth]{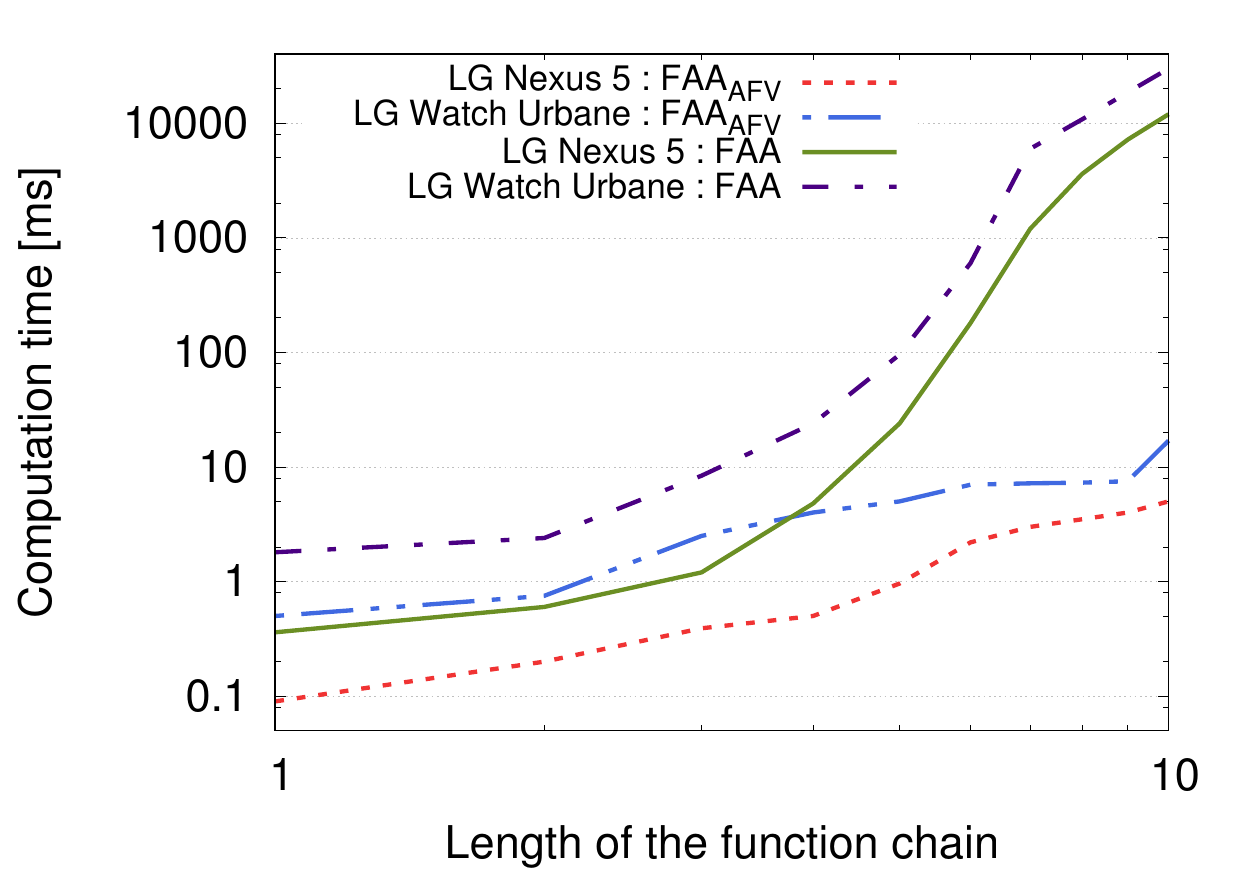}}
	\subfigure[Use case 1]{\label{fig:lifetime_min}\includegraphics[width=0.96\textwidth]{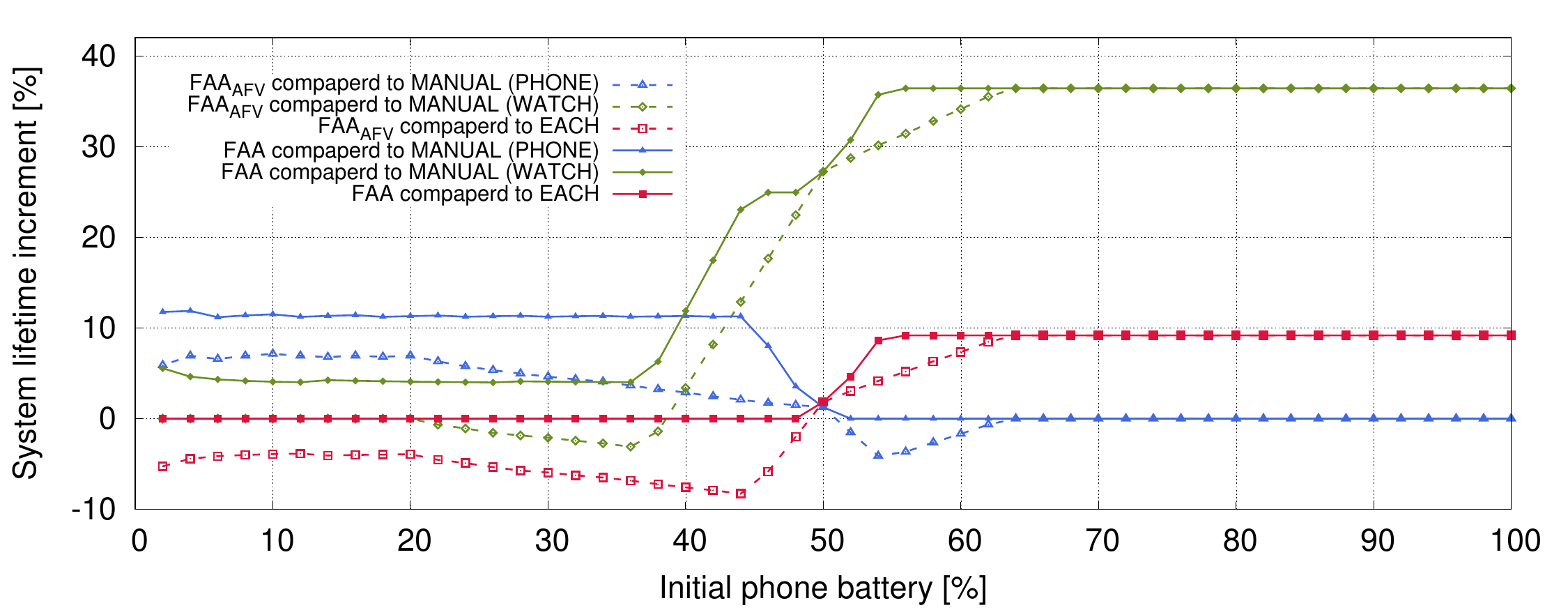}}
	\caption{Accuracy, Efficiency and Robustness of \textsc{faa} Algorithm}
	\label{fig:Evaluation} 
\end{figure}

We measure the percentage increment of the system lifetime when using \textsc{faa} (Algorithm 1) compared to the four existing approaches. Figure \ref{fig:optimscaling} shows the percentage system lifetime increment by using \textsc{faa} when the number of functions in the function chain is increased. As being a heuristic solution method, \textsc{faa} provides a suboptimal solution. However, the error of this heuristic solution to the OPTIMAL solution is minimal where it has less than 5\% average error even when the number of functions in the chain is 10. On the other hand, the system uptime increment becomes much significant compared to MANUAL, EACH, and $FAA_{AFV}$, as the length of the function chain is increased. This is due to that the \textsc{faa} optimizes the allocation of more functions when the length of the chain is increased. This behavior is even applicable when the number of function chains are increased, where more function requests are optimized by \textsc{faa}. In particular, \textsc{faa} improves system lifetime approximately 50\% compared to non-collaborative function allocation such as MANUAL and EACH methods and improves by 40\% compared to collaborative function allocation such as $FAA_{AFV}$.
%

%


As mentioned in Section \ref{sec:Heu}, the heuristic methods of solutions for the MILP problems does not guarantee the efficient solution methods. Therefore, we measure the computation time for the \textsc{faa} algorithm as a measure for the efficiency of the algorithm, and the results are shown in Figure \ref{execution}. Also, we compared the computation time with the $FAA_{AFV}$. Both of these algorithms were developed in Android as wearable/mobile applications and measured the time usage of calculations. We use Nexus5 as the smartphone example and LG Watch Urbane as smartwatch example. As per the results, even when the number of functions per device is 10, \textsc{faa} provides near optimal solutions within 1 second (note the log axes), where the exact method (brute force search) takes few hours to calculate the optimal solution. Moreover, the \textsc{faa} algorithm execution time is comparable with the algorithm execution time of $FAA_{AFV}$, but provides higher quality solutions than $FAA_{AFV}$. 

As per the above results, a PAN consists of three devices and when the number of functions per device is changed from 1 to 10, a minimal error with the optimal solution is maintained by the \textsc{faa} algorithm (cf. Figure \ref{fig:optimscaling}). Also, these sub optimal results are provided within 1 second for all the values of the number of functions (cf. Figure \ref{execution}). These results express the robustness of the proposed \textsc{faa} algorithm for realistic use case scenarios.


\subsubsection{Emulating multiple use case scenarios}

  \begin{table}[ht]
	\caption{Energy cost associated to each function }
	
	\centering
	\scriptsize
	\begin{tabular}{|l|c|c|c|}
		\hline
		\multirow{2}{*}{\textbf{Function}} & \multicolumn{3}{|c|}{\textbf{Energy cost}} \\
		\cline{2-4}
		& \textbf{Phone} & \textbf{Watch} & \textbf{Glasses} \\
		\hline\hline
		
		\multicolumn{4}{|l|}{\textbf{Connectivity (Data Transferring [mW] - Idle [mJ])}}\\\hline
		Bluetooth& 520 - 605  & 180 - 190.3 & 164 - 191.85 \\\hline
		WiFi& 790 - 66 & 480 - 50 & 460 - 52 \\\hline\hline
		\multicolumn{4}{|l|}{\textbf{Processing [mW]}}\\\hline
		Encoding&1400& 860 &900\\\hline\hline
		\multicolumn{4}{|l|}{\textbf{Sensing [mW]}}\\\hline
		Accelerometer-FASTEST & 77.7 & 164 & 153.4 \\\hline
		Video Capturing \cite{LiKamWa2014} & - & - & 2963 \\\hline
		GPS Location \cite{Carroll2010} & 166 & 148 & 155 \\\hline
		Step Count & - & 5 & -\\\hline
		
		
	\end{tabular}
	\label{tab:functionscosts}
\end{table}

Next, a use case scenario with real energy cost values is emulated and present the accuracy and robustness of \textsc{faa}. We consider a PAN that consists of two devices (smartphone and smartwatch), and applications installed in each of the device request accelerometer data in FASTEST speed for the indoor navigation. The application updates in every 30 seconds, and therefore, the data is processed by the applications in every 30 seconds. We modelled this scenario as two requests for the function of \textit{`accelerator data in FASTEST speed'} are made by two devices in the PAN that required to receive data in every 30 seconds. The power requirement specified in Table \ref{tab:functionscosts} are used for the calculations and the smartphone and smartwatch have 2300mAh and 410mAh battery capacities respectively.

We implemented \textsc{faa} and also the algorithm proposed in \cite{Kolamunna2016}. Figure \ref{fig:lifetime_min} shows the system lifetime increment of using each \textsc{faa} algorithm, compared to different methods of existing function allocation, i.e. EACH and MANUAL (smartphone/ smartwatch). The initial battery percentage of the smartphone is changed along with the experiment, providing different use case scenarios. As per the results, \textsc{faa} algorithm always selects the optimal function allocation at all the values of initial phone battery percentages. However, $FAA_{AFV}$ is not always being the optimal, as the selection of $FAA_{AFV}$ decreases the system lifetime compared to some of the common function allocation method (EACH, MANUAL) for certain initial phone battery conditions. The results show that the \textsc{faa} provides accurate results at each different status of the devices at the beginning of the experiment, i.e. different levels of initial battery status.  

\subsection{Evaluation with real world scenarios}

We consider a real world scenario that consists of five devices as shown in Figure \ref{fig:pwn} and they are already inter connected. It has three Tier 1 devices (smartphone, smartwatch, smartglasses), one Tier 2 device (smart-sole) that has the GPS sensing capability, and also a (laptop computer) that has the highest computational capabilities. However, the laptop computer is not always attached with the user, and therefore, we refer to it as an extended PAN device.
Resource hungry gaming application in the smartglasses involves in video streaming. It captures the video of the surrounding, adds the GPS location data, encrypts data and streams to another external party. The smartwatch is running a fitness tracking application that requires GPS location and step count at the same time. Also, a navigation application is running in the smartphone that requires GPS location. Here, the user/developer has specified preferences to run the \textit{`Capture Video'} function in the smartglasses, and the \textit{`Step Count'} function in the smartwatch. Therefore, these two functions are assigned accordingly at the beginning of the \textsc{faa}. The remaining function requests are allocated as per the \textsc{faa}. The specified energy values in Table \ref{tab:functionscosts} is used to emulate the real world scenarios. The smartphone, smartwatch and smartglasses has 2300mAh, 410mAh and 570mAh battery capacities respectively. In order to calculate the default energy usage, we consider the smartphone can retain for 2 days with the default energy drainage, smartwatch and smartglasses retain one day. These were obtained from the specifications for real devices.

\begin{figure}[tb]
	\centering 
	\includegraphics[width=1\textwidth]{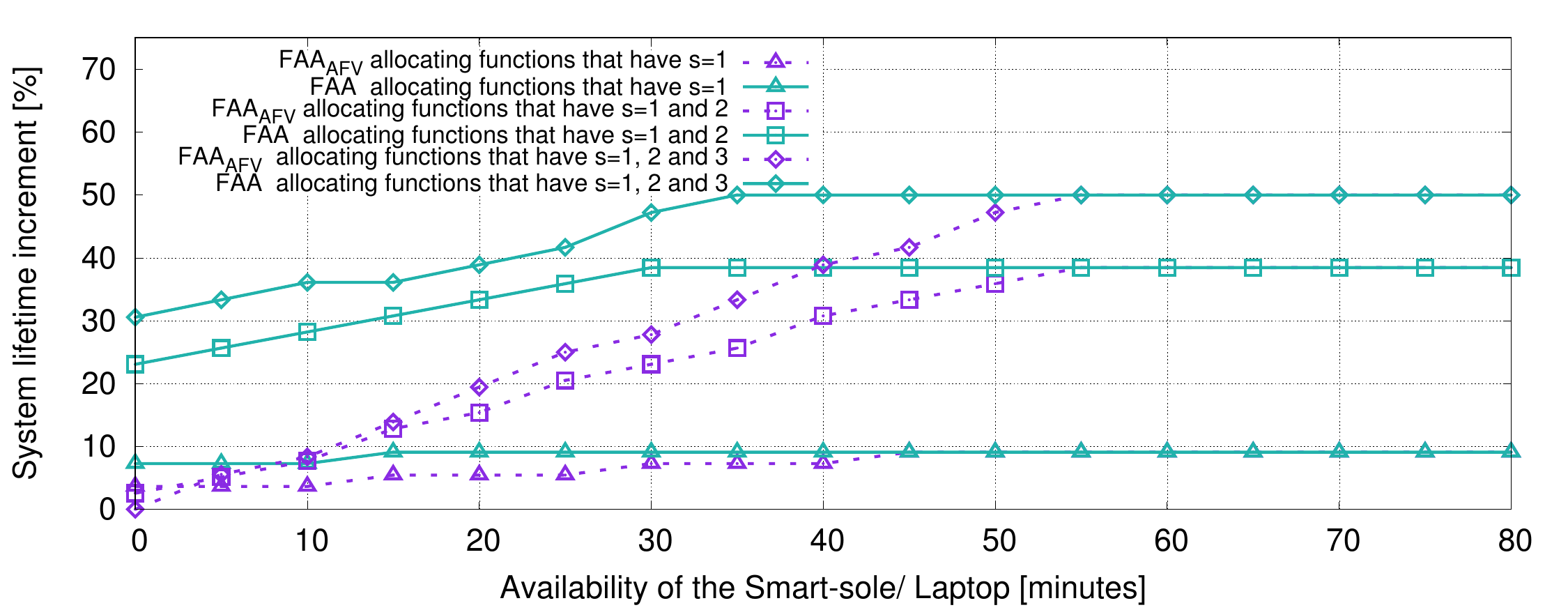}
	\caption{Function Chaining.}
	\label{fig:ratio_uptime}
\end{figure}

Figure \ref{fig:ratio_uptime} shows the percentage of system lifetime increment compared to the current default method of function allocation (i.e. the device itself is performing all the functions), against the availability of the Tier 2 device (smart-soles) and the extended PAN device (laptop computer). When the tier 2 device and the extended PAN device are available for a longer time period, the system uptime increment inclines. However, after a certain time, the system lifetime can not be further increased, and that is where each graph is plateaued. This is because that, even all the processing and sensing are fully allocated to the laptop computer and the smart-sole, while allocating the minimum possible amount of workload to the Tier 1 devices, the Tier 1 devices will get their batteries drained out and limit the further increment of the system lifetime.
 
Moreover, same as in Section \ref{Eval}, it can be observed that the percentage system lifetime increment is much more significant compared to the default function allocation when the length of the function chain is increased. Also, the \textsc{faa} algorithm proposed in this paper outperforms the $FAA_{AFV}$ at each scenario of availability of the devices.


%
%
%
%

\section{Conclusion}
\label{conclusion}

Smart wearable devices is increasing exponentially and are becoming pervasive.  We are also witnessing a wide variety of fascinating new applications and services, such as mixed reality,  that leverage the capabilities of these wearables, but also massively resource hungry. However, one of the major constraints of wearables is the battery capacity, and as a result, this leads to poor user experience as the application lifetimes become limited.  Moreover, the current non-efficient resources utilization exacerbates  this problem. It is clear that one of the most practical solutions to the problem is to leverage the resources of multiple devices on the personal area network. However, this requires efficient coordination of the pool of resources that are available and the orchestration of their use.

To the best of our knowledge, this paper presented the first proposal for application function orchestration, and showed that it is possible to maximize the overall wearable system lifetime using a heuristic algorithm of function allocation. The percentage system lifetime increases with the length and the number of function chains as it increases the number of optimized function allocations. For a particular use case of 3 devices and 10 function types, the proposed heuristic method of function allocation increases the system lifetime by $\sim$50\% compared to the existing common non-collaborative function allocation methods, i.e. manual allocation and allocation to all devices, and by $\sim$40\% compared to the most related collaborative function allocation method proposed in the literature. However, the heuristic solution methods used in the proposed algorithm maintains a minimal error with the optimal solutions while providing a significant improvements in the time complexity and efficiency. We also demonstrated the practical feasibility of executing the proposed algorithm on real devices in near real time by implementing it on a smartphone and a smartwatch. 

In our future work, we intend to extend our implementation to develop a cross-device function allocation framework that can be utilized by application developers to take advantage of the power of common functionalities available on the PAN.

\bibliographystyle{ACM-Reference-Format}
\bibliography{sample-bibliography}

\appendix

\section{Brute Force Search}
\label{BFS}

\textit{Example :}
Let's consider an example use case where there are two devices (i.e., Smartphone-P and Smartwatch-W), and there are two functionality requests coming from each device for accelerometer data and gyroscope data. Applications installed in each device is requesting data in every 1 minute. 

In this method, we consider all the possible assignments of the functionalities and find the maximum of the ON time of any of the devices in each assignment. Then we find the assignment that has the maximum from this maximum ON time. We calculate the computational complexity of brute force search.
\begin{itemize}
	\item Number of devices -~$\mathbf{n}$
	\item Number of function types -~$\mathbf{x}$
	\item Number of requests for each function type -~$\mathbf{r}$
	\item Number of devices that has the function type $\mathbf{f}$ -~$\mathbf{n_f}$
	\item Number of requests from the function type $\mathbf{f}$ -~$\mathbf{r_f}$
	\item[] Number of Combinations ($\mathbf{k}$) = $\mathbf{n_1^{r_1}\times n_2^{r_2}\times...\times n_x^{r_x}}$
	\item[] Worst case \tabto{1cm}$\mathbf{n_1=n_2=...=n_x=n}$\\
	\tabto{1cm}$\mathbf{r_1=r_2=...=r_x=r}$\\
	
	\tabto{1cm} $\mathbf{k={(n^{rx})}}$
\end{itemize}


\begin{itemize}
	\item Calculating each entry in the table and check whether the entities satisfies the each device's requirement = $\mathbf{\Theta(n\cdot k)}$ = $\mathbf{\Theta(n^{(rx+1)})}$
	\item Finding the minimum in one combination = $\Theta(n)$
	\item Finding the minimum in each combination = $\mathbf{\Theta(n\cdot k)}$ = $\mathbf{\Theta(n^{(rx+1)})}$
	\item Finding the minimum among all the minimum values = $\mathbf{\Theta(k)}$ = $\mathbf{\Theta(n^{rx})}$
	\item[] Total Complexity \tabto{1cm}= $\mathbf{\Theta(2\cdot n^{(rx+1)}+n^{rx})}$
\end{itemize}

As expected, the calculated complexity of the brute force method is in exponential time.

\section{Branch and Bound Method}
\label{BB}
These algorithms are nonheuristic, in the sense that the sub optimal solutions are eliminated only after making sure that they do not lead to the optimal solution. 
Let's consider an example with real cost values. Table \ref{tab1} shows the real energy values per minute. We consider remaining battery is 20\% in the smartphone (which has $2300mAh$ for the 100\% battery capacity) and 70\% in the smartwatch (which has $410mAh$ for the 100\% battery capacity). Moreover, we consider smartphone and smartwatch are consuming $150mJ$ and $70mJ$ per minute respectively. In this example scenario, we neglected the idle energy for the communication which are $600mJ$ in smartphone and $190mJ$ in smartwatch.

There are different functionality requests coming from the devices. These different types of requests can not be mapped to the other types of functions. As an example, accelerometer functionality request can not be fulfilled by running the gyroscope. 

As a solution, we can try and consider individual optimization problems for different functionalities and then combine the results to get the final results. However, this particular optimization objective does not allow to go for that type of solutions as the selections of one type of functionality directly influences the solutions of other types of functionalities.  

Therefore, in order to avoid these types of difficulties, we alter the consideration of the first selection of the functionalities. Basically, we group the functionalities just for the first selection in a way that one group contains all the requested functionality types. We consider every possibility of combinations. Once we select the best combination of functions as per the objective, we check with the additions and removal of each and every other functionality from the next step.

\begin{itemize}
	\item [] \textit{\textbf{Cost Value :}} Here we consider the inverse of the maximum of the ON time of any of the devices. At each step we consider the maximum ON time of any of the devices. This cost value has two bounds which are the \textit{Upper Bound} and \textit{Lower Bound}.
	\item [] \textit{\textbf{Upper Bound :}} The local minimum of the cost value. This value is calculated by connecting request only to the opened functions.
	\item [] \textit{\textbf{Lower Bound :}} The local optimum solution that would lead from the particular solution is always higher than the lower bound. This value is obtained by connecting requests to the opened and unknown functions, but not to the closed functions.
	\item [] \textit{\textbf{Pruning Rules :}} If the \textit{Upper Bound} is larger than the parent node, the branch is pruned.
\end{itemize}

\begin{table}[h]
	\caption{Energy costs per minute. \textit{(Phone's energy usage, Watch's energy usage)}}
	\centering
	\begin{tabular}{|c||c|c|c|c|c|}
		
		\hline
		$d_i,v_x$&$f$ & $r_{p,acc}$ & $r_{p,gyro}$ & $r_{w,acc}$ & $r_{w,gyro}$ \\ 
		\hline\hline
		$X_1$~~~$p,acc$  &(300,0)& (0,0) & ($\infty$,$\infty$) & (142,36) & ($\infty$,$\infty$)  \\ 
		\hline
		$X_2$~~~$p,gyro$  &(660,0)& ($\infty$,$\infty$)  &  (0,0) & ($\infty$,$\infty$)  &  (142,36)\\ 
		\hline
		$X_3$~~~$w,acc$  &(0,600)& (142,36) & ($\infty$,$\infty$)  &  (0,0) & ($\infty$,$\infty$)  \\ 
		\hline
		$X_4$~~~$w,gyro$  &(0,960)& ($\infty$,$\infty$)  & (142,36) &($\infty$,$\infty$)   &(0,0)  \\ 
		\hline
	\end{tabular}
	\label{tab1}
\end{table}

\begin{figure}[h]
	\centering 
	\includegraphics[width=0.9\textwidth]{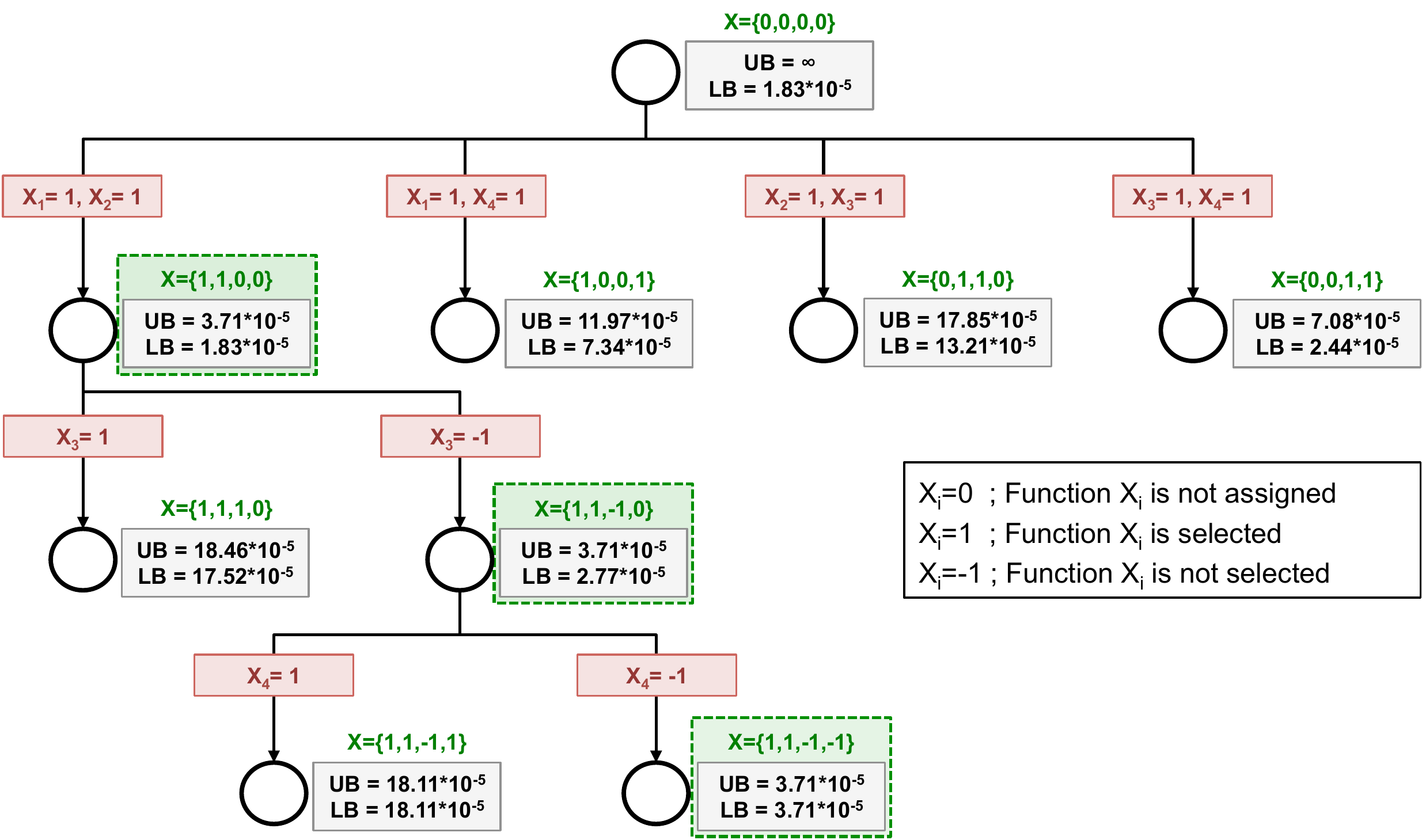}
	\caption{An example for Branch and Bound method}
	\label{fig1}
\end{figure}

In contrast to brute force search, in this method, we do not consider every possible combination of the functions allocation in this method as we prune some branches once we confirm that they do not lead to the optimum solution.

Next we calculate the computational complexity of this method for this particular case. Let's assume the $\mathbf{n}$, $\mathbf{x}$ and $\mathbf{r}$ has the same meanings as in the previous sub section.

First, we consider the computational complexity at the first branching. 
\begin{itemize}
	\item Taking the possible combinations for the first branching. \\\tabto{1cm}= $\mathbf{O(n^x)}$
	\item Calculating UB for each combination. 
	\begin{itemize} 
		\item As all the requests for $f_i$ are assigned to the function $v_i$, there are no selected assignments. Calculate $\frac{1}{T_i}$ for all the devices, check whether the $\frac{1}{T_i} \le \frac{1}{T^*_i}$, and select the minimum. \\\tabto{1cm}= $\mathbf{\Theta(3n)}$
	\end{itemize}
	\item Calculating the LB for each combination.
	\begin{itemize} 
		\item Consider each device separately, and for each request try to get the assignment that requires least energy from the considered device. \\\tabto{1cm}= $\mathbf{\Theta(n\cdot x\cdot r \cdot x \cdot n)}$ \\\tabto{1cm}= $\mathbf{\Theta(n^2x^2r)}$
		\item Select the min of $\frac{1}{T_i}$ \\\tabto{1cm}= $\mathbf{\Theta(n)}$
	\end{itemize}
	\item Total complexity of first branching \\\tabto{1cm}= $\mathbf{\Theta(n^x+2n+n+n^2x^2r+n)}$ \\\tabto{1cm}= $\mathbf{O(n^x+n^2x^2r)}$
\end{itemize}

Next, we calculate the computational complexity during rest of the branching. The selection of the next function to be opened is decided by the fact that the most popular function during the lower bound calculation and that is not opened yet. 
\begin{itemize}
	\item Calculating each value of the next branches from the initial selection \\\tabto{1cm}= $\mathbf{\Theta(2+2^2+2^3+...+2^{[x\cdot n -x]})}$ \\\tabto{1cm}= $\mathbf{\Theta(2+2^2+2^3+...+2^{x[n-1])}}$
	\item Calculating UB for each combination 
	\begin{itemize} 
		\item Consider each device separately, and for each request try to get the assignment that requires least energy from the particular device \\\tabto{1cm}= $\mathbf{O(n\cdot x\cdot r \cdot x \cdot n)}$ \\\tabto{1cm}= $\mathbf{O(n^2x^2r)}$
		\item Check whether the $\frac{1}{T_i} \le \frac{1}{T^*_i}$ \\\tabto{1cm}= $\mathbf{\Theta(n)}$
		\item Select the min of $\frac{1}{T_i}$ \\\tabto{1cm}= $\mathbf{\Theta(n)}$
	\end{itemize}
	\item Calculating the LB for each combination
	\begin{itemize} 
		\item Consider each device separately, and for each request try to get the assignment that requires least energy from the particular device \\\tabto{1cm}= $\mathbf{O(n\cdot x\cdot r \cdot x \cdot n)}$ \\\tabto{1cm}= $\mathbf{O(n^2x^2r)}$
		\item Select the min of $\frac{1}{T_i}$ \\\tabto{1cm}= $\mathbf{\Theta(n)}$
	\end{itemize}
	\item Total complexity of next branching\\ \tabto{1cm}= $\mathbf{O \Big(\big(2n^2x^2r+3n\big)*\big(n^x+2+2^2+2^3+...+2^{x[n-1]}\big)\Big )}$
	\tabto{1cm}= $\mathbf{O \Big (\big(2n^2x^2r+3n\big)*\big(2^{x[n-1]+1}-2\big)\Big)}$
	\tabto{1cm}= $\mathbf{O  \big(n^2x^2r2^{x[n-1]+2}+3n2^{x[n-1]+1}-4n^2x^2r-6n\big)}$
\end{itemize}
The overall complexity is the aggregation of the two steps. However, as expected, this method also provides the exponential complexity.
\begin{itemize}
	\item[] Total Complexity
	\tabto{1cm}= $\mathbf{O  \big(n^x+n^2x^2r+n^2x^2r2^{x[n-1]+2}+3n2^{x[n-1]+1}-4n^2x^2r-6n\big)}$
\end{itemize}

\end{document}